\documentclass[letterpaper,times]{IONconf}

\usepackage{amsmath, amssymb} 
\usepackage{graphicx}
\usepackage{url}
\usepackage{natbib}
\usepackage[hidelinks]{hyperref}
\usepackage{array}
\usepackage{multirow}

\usepackage{cleveref}
\Crefname{lstlisting}{Listing}{Listings} 
\Crefname{lstlisting}{Listing}{Listings} 

\usepackage{listings}
\lstdefinestyle{pseudopy}{
  language=Python,
  keywordstyle=\color{blue},
  commentstyle=\color{gray},
  stringstyle=\color{red},
  showstringspaces=false,
  frame=single,
  breaklines=true,
  numbers=left,
}

\usepackage{fontawesome5}
\usepackage{tikz}
\usetikzlibrary{positioning}
\usetikzlibrary{decorations,decorations.pathmorphing,decorations.text, decorations.pathreplacing}
\usetikzlibrary{shapes.geometric}

\newcommand\mperiod[1][\rlap]{#1{\;.}}
\newcommand\mcomma[1][\rlap]{#1{\;}} 
\newcommand\msemi[1][\rlap]{#1{\;;}}

\title{World’s First Authenticated Satellite Pseudorange from Orbit}
\author{
    Jason~Anderson, \textit{Xona Space Systems}
    }

\date{11 September 2025}

\begin{document}

\maketitle

\section*{biography}

\biography{Jason Anderson}{
    works at Xona as a Senior Security Engineer and a Stanford GPS Lab Alumnus.
    For his doctoral thesis, Designing Cryptography Systems for GNSS Data and Ranging Authentication, he won the 2025 ION Parkinson Award, the 2025 RTCA Jackson Award, and the Bauhaus Prize.
    Before Stanford, Jason received a Mechanical and Nuclear Engineering degree from UC Berkeley and has worked at SpaceX and General Atomics.
}

\section*{Abstract}
    Cryptographic Ranging Authentication is here!
    We present initial results on the Pulsar authenticated ranging service broadcast from space with Pulsar-0 utilizing a recording taken at Xona headquarters in Burlingame, CA.
    No assumptions pertaining to the ownership or leakage of encryption keys are required.
    This work discusses the Pulsar watermark design and security analysis.
    We derive the Pulsar watermark's probabilities of missed detection and false alarm, and we discuss the required receiver processing needed to utilize the Pulsar watermark.
    We present validation results of the Pulsar watermark utilizing the transmissions from orbit.
    Lastly, we provide results that demonstrate the spoofing detection efficacy with a spoofing scenario that incorporates the authentic transmissions from orbit.
    Because we make no assumption about the leakage of symmetric encryption keys, this work provides mathematical justification of the watermark's security, and our July 2025 transmissions from orbit, we claim the world's first authenticated satellite pseudorange from orbit.

\section{Introduction}

Pulsar is a Low-Earth-Orbit (LEO) satellite navigation service provided by Xona Space Systems.
Security is core a feature of Pulsar, providing cryptographic data authentication, cryptographic ranging authentication via a watermark, and User-specific Dynamic Authenticated Encryption~\citep{anderson2025DAE}.
This work introduces and provides first results for Pulsar's cryptographic ranging authentication service.

Cryptographic ranging authentication requires that a navigation ranging signal perturb its ranging codes in a cryptographic fashion.
Because these cryptographic perturbations are infeasible for an adversary to predict before their broadcast, they can be used to establish trust in the derived pseudoranges from the navigation service.
These perturbations are designed to be small to limit the effects on tracking and to those receivers that do not require ranging authentication.
Cryptographic ranging via watermarking has been studied and proposed many times among navigation researchers~\citep{Anderson2017,fernandez2024galileo}.
So far, results have pertained to theoretic signals.

In June 2025, Xona Space Systems launched Pulsar-0, the first satellite among the Pulsar satellite navigation service constellation.
On July 7th, 2025, while conducting Pulsar-0's commissioning, Xona broadcast watermarked ranging codes.
Thereafter, the ranges were authenticated utilizing a software-defined radio.
Therefore, Xona Space Systems claims the world's first authenticated pseudorange from orbit and is pleased to offer customer demos and commercial service in the near future.

\begin{figure}
    \centering
    \begin{minipage}[t]{0.48\linewidth}
        \centering
        \includegraphics[width=0.9\linewidth]{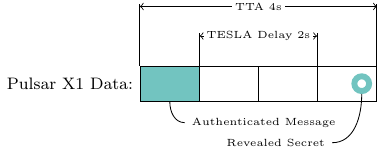}
        \caption{
            A conceptual diagram listing out the components that make up TTA.
            TTA means the time of the first bit of authenticated information through the time of the last needed bit required to authenticate that information.
            Pulsar X1 messages are transmitted each second.
            The TTA components include one second for the authenticated information (constellation wide data and ranging), the required TESLA delay (configurable, but targeting two seconds), and one second to receive a message containing the needed authentication secret to authenticate information constellation wide.
        }
        \label{fig:conceptualTTA}
    \end{minipage}
    \hfill
    \begin{minipage}[t]{0.48\linewidth}
        \centering
        \includegraphics[width=\linewidth]{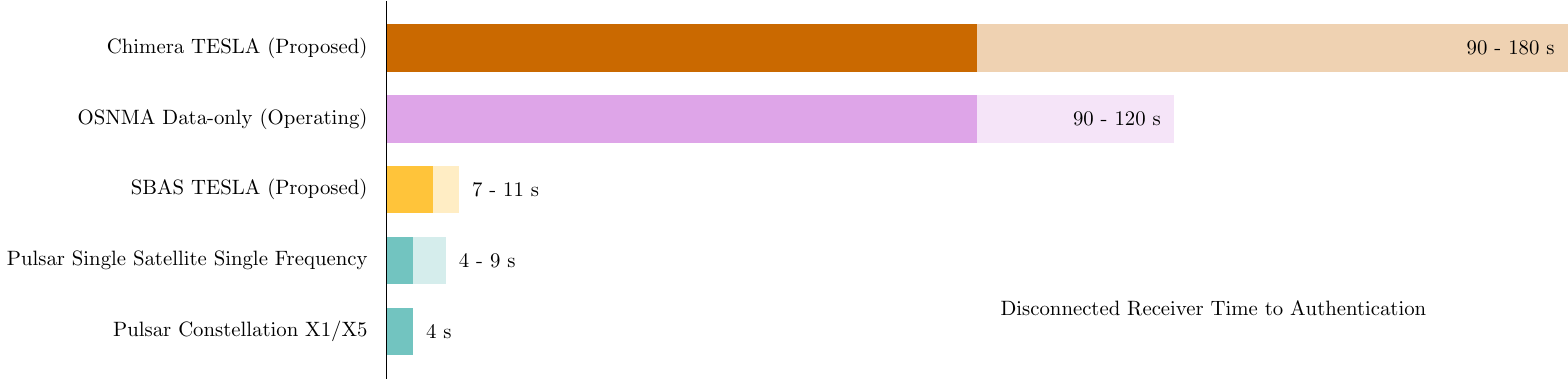}
        \caption{
            A comparison of the TTA of several operating and proposed authenticated PNT signals~\citep{chimeraicdTESLA,Galicd,improvOSNMA,dennis2024sbas}.
            Pulsar provides demonstrably faster TTA for both data and ranging authentication without requiring a network connection.
        }
        \label{fig:compareTTA}
    \end{minipage}
\end{figure}

The Pulsar X1 signal broadcasts on the L1 band and utilizes 1023-chip ranging codes.
While Pulsar also transmits an L5 signal called X5, this work focuses exclusively on experiments with the Pulsar X1 service on the L1 band.
The Pulsar X1 Watermark is a Combinatorial Watermark where exactly 21 chips among the 1023 are inverted.
Pulsar's X1 Watermark follows the design procedure from \citet{anderson2024gnsscrypto}.
These 21 chips are pseudorandomly derived from cryptographic secrets released in the Pulsar data channel and are constructed to provably assure that no adversary can predict them before broadcast.
The watermark is designed to enable a receiver to complete an authentication determination utilizing 1 second of ranging IQ data.
Once the entire constellation of satellites is operating, the Pulsar Ranging authentication is targeting a time-to-authentication of four seconds for all dual frequency receivers without the need of a network connection during receiver operation.
\Cref{fig:conceptualTTA} provides a conceptual diagram of the components the make Pulsar's time to authentication (TTA) of four seconds.
\Cref{fig:compareTTA} provides a comparison of Pulsar's TTA to those of other proposed or operating services.

\begin{figure}
    \centering
    \begin{minipage}[t]{0.48\linewidth}
        \centering
        \includegraphics[height=0.6\linewidth]{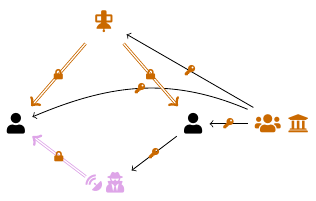}
        \caption{
            A conceptual diagram of a security vulnerability inherent to signals that only utilize encryption cryptography.
            With an encryption-only service, encryption keys are distributed to the satellite constellation and all authorized users.
            If an incompetent or dishonest receiver leaks the encryption keys to a spoofer, then that spoofer can forge signals to all users.
            To mitigate this risk, receivers must be vetted and include tamper-resistant memory, and rekeying requires frequent human touchpoints.
            The keys and single-line arrows represent the transfer of encryption keys, either via network or via personnel.
            The double-line arrows represent PNT signals, and the locks represent how the signal is encrypted to provide confidentiality.
        }
        \label{fig:worseSecurity}
    \end{minipage}
    \hfill
    \begin{minipage}[t]{0.48\linewidth}
        \centering
        \includegraphics[height=0.6\linewidth]{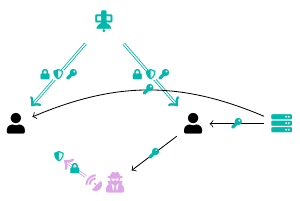}
        \caption{
            A conceptual diagram of the Pulsar authentication information distribution.
            Authentication is completed with public-key (whose secrets are never distributed) and TESLA (whose secrets are revealed at the correct time to provide authentication) cryptography.
            Even if a user leaks its encryption keys to a spoofer, the spoofer cannot forge the authentication cryptography before broadcast.
            The keys and single-line arrows represent the transfer of encryption keys, either via network or over the air on the Pulsar data stream.
            The double-line arrows represent PNT signals, and the shields represent how the signal is protected with authentication cryptography regardless of whether the specific messages are encrypted or not encrypted.
        }
        \label{fig:betterSecurity}
    \end{minipage}
\end{figure}

At present, there are several GNSS systems that provide an encrypted service that purport to provide authenticated ranges.
However, these systems prohibit access to non-military and non-government users and contain a security vulnerability related to the encryption key methodology.
With an encrypted service, the GNSS provider and groups of authorized receivers share signal encryption keys.
Anyone with the signal encryption keys can generate the encrypted signals.
This enables any authorized user or any user with a leaked encryption key to spoof the encrypted signals.
\Cref{fig:worseSecurity} provides conceptual diagrams of the security vulnerability of encrypted services that purport authentication, and \Cref{fig:betterSecurity} provides a conceptual diagram of Pulsar's authentication service showing that it does not contain this vulnerability.

Since Pulsar will provide a commercial satellite navigation signal, if Pulsar were to adopt this symmetric encryption design for the purpose of claiming an authenticated service, anyone could purchase the encryption keys and then spoof the Pulsar signal.
Therefore, Pulsar cannot adopt the authentication paradigm present in today's encrypted signals.
Instead, only Pulsar has the cryptographic secrets needed to produce the Pulsar ranging service.
This means that the Pulsar signal elevates its authentication standard to that of standard cryptographic authentication.
We note that the Pulsar watermark is still vulnerable to the Security Code and Estimation Replay (SCER) adversary, which is fundamental to any ranging service~\citep{Psiaki2016}.

In the following sections we discuss the Pulsar Watermark and the results needed to claim the world's first authenticated pseudorange from orbit.
\Cref{sec:notation} introduces notation, this work's adversarial model, and the signal processing needed to utilize the Pulsar X1 Watermark.
\Cref{sec:wm-design} discusses design considerations for the Pulsar Watermark, and \Cref{sec:wm-sec} derives the Pulsar X1 Watermark security.
Utilizing a receiver recording covering a Pulsar-0 orbit pass over Burlingame, CA, \Cref{sec:resultsFromOrbit} provides watermark authentication results constituting the worlds first authenticated pseudorange from orbit.
From the same data, \Cref{sec:spoofExp} provides results from a quick spoofing experiment to demonstrate the Pulsar Watermark's efficacy.

\section{Notation, Adversarial Model, and Signal Processing}\label{sec:notation}

We refer to \citet{anderson2024gnsscrypto} for the complete derivations of the math from this work.
For the readers convenience, \Cref{tab:notation} provides a notation table of the Pulsar X1 Watermark.
The Pulsar X1 signal is transmitted on the GNSS L1 band, and like several GNSS signals presently transmitting, the X1 ranging codes are 1023 chips long.

\begin{table}
    \centering
    \begin{tabular}{|p{0.20\linewidth}|p{0.75\linewidth}|} \hline
        \textbf{Notation} & \textbf{Description} \\ \hline 
        $n = 1023$ & The number of chips over a single watermarked ranging code covering one millisecond. \\
        $r = 21$ & The number of inverted chips per $n$. \\
        $W = 1000$ & The number of individual watermarked ranging codes that must be considered together for a single authentication determination. \\ 
        $F = 2.046$ MHz & The radio's measurement frequency. This value corresponds to the Nyquist frequency, forming a worst-cast sampling rate appropriate for this work's security analysis. As radios increase their sampling frequency, the security afforded gets better. \\
        $T = 1$ ms & The length of time over which a single watermark applies. \\
        $F \cdot T = 2046$ & The number of samples per ranging code. \\
        $C/N_0 = 30$ dB-Hz & The carrier-to-noise Ratio. This value corresponds to a worst-case assumption appropriate for this work's security analysis. As radios observe better $C/N_0$, the security afforded will gets better. \\
        $P = 1$ & The signal power immediately preceding correlation, set to unity without loss of generality. \\
        $\sigma^2 = \frac{P}{C/N_0}\frac F2$ & The pilot noise power immediately preceding correlation. \\ \hline
    \end{tabular}
    \caption{Pulsar Watermark And Receiver Processing Model Notation Table}
    \label{tab:notation}
\end{table}

The Pulsar X1 Watermark is a Combinatorial Watermark~\citep{anderson2024gnsscrypto}.
Every ranging code will have a fixed number of inversions that are pseudorandomly selected to comply with a bit-commitment authentication protocol.
While the Pulsar X1 Watermark inverted-chip count is configurable, we expect that Pulsar will always invert exactly 21 chips per 1023-chip ranging code.
The 21-chip election requires that the receiver aggregate 1000 watermarked ranging codes to a make an authentication determination (corresponding to one second).
The Pulsar Watermark integrates with the data authentication protocol on the data channel, which is based on TESLA like Galileo's OSNMA~\citep{Galicd}.

The adversarial model of this work is the non-SCER adversary, which does not engage in SCER attacks~\citep{Psiaki2016}.
This means the mathematics of this work assumes that the adversary will not attempt to observe the watermark before submitting a spoofed watermarked ranging code to a victim receiver.
Instead, the adversary will spoof by transmitting ranging codes with randomly drawn inverted chips.
The adversary may elect to invert any number of chips $s$ among the $n$ before submitting a spoofed signal to its victim.
Analysis of the Pulsar watermark under SCER spoofers is reserved for future work.

To complete our watermark security model, we form a conservative radio model.
The receiver should have a sampling rate that is at least the Nyquist frequency.
Moreover, the receiver should not attempt to authenticate the signal if its measured $C/N_0$ is less than 30 dB-Hz.
These assumptions form a conservative worst-case appropriate for the security analysis of this work.
We expect receivers should always operate with better radios and in better conditions, meaning that the actual security for a receiver should always be better than what is derived herein.
We note that if this conservative radio model is later found to not be conservative enough, the security level requirement can be maintained by increasing $W$ or $r$, increasing the watermark aggregation or the inverted chip count, respectively.

\begin{figure}
\centering
\includegraphics[width=0.75\linewidth]{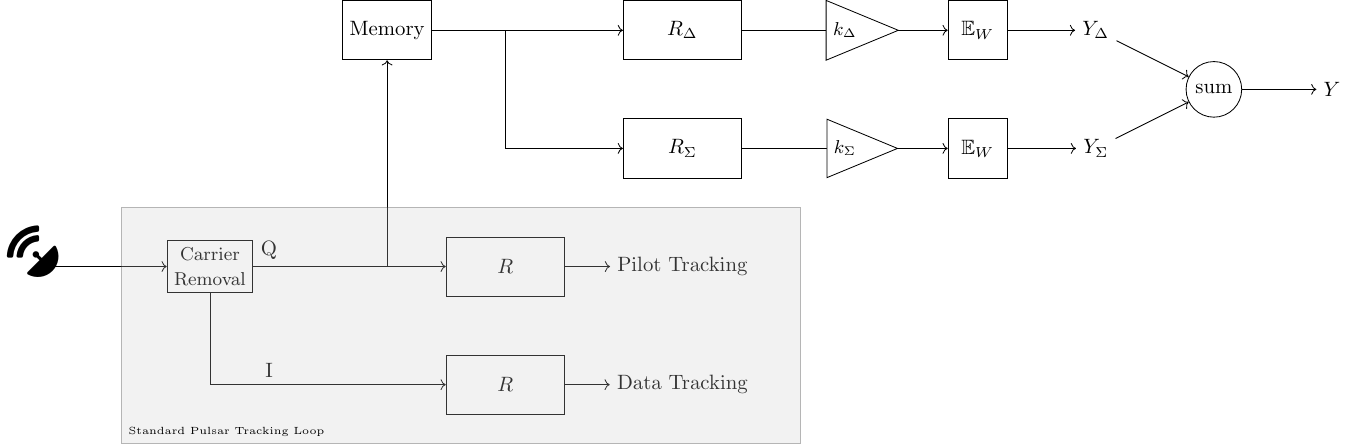}
\caption{A conceptual diagram of the watermark signal processing within a Pulsar receiver.}
\label{fig:concept-wm-radio}
\end{figure}

\Cref{fig:concept-wm-radio} provides a conceptual diagram of the receiver's required signal processing to make an authenticity determination.
\Cref{tab:concept-wm-radio} lists the definitions of each of the variables within \Cref{fig:concept-wm-radio}.
To determine the authenticity of the signal, a receiver must first track the signal without knowledge of the watermark and store the IQ samples and tracking variables (e.g., Phase, Doppler) for later.
Because $r=21$, this amounts to a $\frac{n - 2r}{n} = -0.364$ dB degradation of the nominal signal.
Later, Pulsar will transmit information that reveals the watermark.
At that time, the receiver will need to perform authentication checks on those stored IQ samples to determine the signal authenticity.

\begin{table}
    \centering
    \begin{tabular}{|p{0.20\linewidth}|p{0.75\linewidth}|} \hline
        \textbf{Notation} & \textbf{Description} \\ \hline
        $R^w, R$ & The watermarked and unwatermarked replicas for the signal over a single resampled ranging code, respectively. Each is a vector of $-1$ and $1$ with length $FT$. \\
        $R_\Delta = R^w - R$ & The kernel of the $R_\Delta$ matching filter with length $FT$. \\
        $R_\Sigma = R^w + R$ & The kernel of the $R_\Sigma$ matching filter with length $FT$. \\
        $k_\Delta = \frac{1}{2r}\frac{n}{FT}\frac{1}{P}$ \rule{0pt}{2.5ex} & The gain on the $R_\Delta$ filter. \\
        $k_\Sigma = \frac{1}{2(n-r)}\frac{n}{FT}\frac{1}{P}$ \rule{0pt}{2.5ex} & The gain on the $R_\Sigma$ filter. \\
        $Y_\Delta$ & The random variable representing the output from the $R_\Delta$ filter after a $W$ average. \\
        $Y_\Sigma$ &  The random variable representing the output from the $R_\Sigma$ filter after a $W$ average. \\
        $Y$ & The random variable that results from summing of $Y_\Delta$ and $Y_\Sigma$. \\ \hline 
    \end{tabular}
    \caption{Pulsar Watermark Receiver Processing Filter Notation Table}
    \label{tab:concept-wm-radio}
\end{table}

This work derives and discusses the distributions $Y_\Delta$ and $Y_\Sigma$ for the purpose of providing authentication security.
\Cref{tab:wm-dist} lists the relevant distributions and their notations for the derivations contained herein.
\begin{table}
    \centering
    \begin{tabular}{|p{0.3\linewidth}|p{0.65\linewidth}|} \hline
        \textbf{Notation} & \textbf{Description} \\ \hline 
        $N_\Delta \sim \mathcal{N}\left(0, \frac1r \frac{n}{FT} \frac{\sigma^2}{P}\right)$ \rule{0pt}{3ex} & The normal random variable that representing the noise power preceding correlation propagated through the $R_\Delta$ filter. \\
        $N_\Sigma \sim \mathcal{N}\left(0, \frac{1}{n-r} \frac{n}{FT} \frac{\sigma^2}{P}\right)$ \rule{0pt}{3ex} &  The normal random variable that representing the noise power preceding correlation propagated through the $R_\Sigma$ filter. \\
        $N_{\Delta \Sigma W}$ & The normal random variable that results from averaging $W$ sums of $N_\Delta$ and $N_\Sigma$. Its definition is available in \Cref{eq:NDSW}. \\ 
        $s$ & The number of chips the adversary elects to invert when attempting to spoof a watermarked ranging code. \\
        $H \sim \mathcal{H}(n, r, s)$ & The hypergeometric random variable, which specifies the distribution an adversary will guess correctly among the $r$ among the $n$ with $s$ attempts. \\
        $g_\Delta(h) = \frac{1}{2r} (4h - 2r)$ \rule{0pt}{2.5ex} & A linear function that transforms the support of $H$ to $Y_\Delta$. \\
        $g_\Sigma(h) = \frac{1}{2(n-r)} (2n-2r-4s+4h)$ \rule{0pt}{2.5ex} & A linear function that transforms the support of $H$ to $Y_\Sigma$. \\ \hline
    \end{tabular}
    \caption{Pulsar Watermark Statistic Distribution Notation Table}
    \label{tab:wm-dist}
\end{table}

\subsection{CLT Non-SCER Spoofing and Authentic Statistic Distributions}

In this section, we derive the Central Limit Theorem (CLT) approximations for the non-SCER spoofing distributions for statistics $Y_\Delta$ and $Y_\Sigma$.
Applying the CLT is useful with predicting $Y_\Delta$ and $Y_\Sigma$ under spoofing conditions because the receiver will aggregate $W=1000$ identically distributed $Y_\Delta$ and $Y_\Sigma$.
Therefore, we must simply compute the expectation and variance of $Y_\Delta$ and $Y_\Sigma$ under non-SCER spoofing conditions.
The distributions here will be utilized in \Cref{sec:resultsFromOrbit,sec:spoofExp}.

From \citet{anderson2024gnsscrypto}, we have the distribution of $Y_\Delta$ and $Y_\Sigma$ under non-SCER spoofing conditions:
\begin{align}
Y_\Delta \mid \textrm{Non-SCER Spoof} &= g_\Delta(H) + N_\Delta \msemi \label{eq: YD nSCER} \\
Y_\Sigma \mid \textrm{Non-SCER Spoof} &= g_\Sigma(H) + N_\Sigma \mperiod \label{eq: YS nSCER} 
\end{align}
Recall the expectation of hypergeometric distribution $H$:
\begin{align}
    \mathbb{E}\left[H\right] &= \frac{sr}{n} \mperiod
\end{align}
Starting with $Y_\Delta$,
\begin{align}
    \mathbb{E}_W\left[Y_\Delta \mid \text{non-SCER Spoof}\right] &= \mathbb{E}\left[g_\Delta(H) + N_\Delta \right] \msemi \nonumber \\
    \mathbb{E}_W\left[Y_\Delta \mid \text{non-SCER Spoof}\right] &= \mathbb{E}\left[g_\Delta(H) \right] \mperiod \nonumber
\end{align}
Because $g_\Delta$ is a linear function,
\begin{align}
    \mathbb{E}_W\left[Y_\Delta \mid \text{non-SCER Spoof}\right] &= g_\Delta\left(\mathbb{E}\left[H \right]\right) \nonumber \\
    &= g_\Delta\left(\frac{sr}{n}\right) \msemi \nonumber \nonumber \\
    \mathbb{E}_W\left[Y_\Delta \mid \text{non-SCER Spoof}\right] &= \frac{2s}{n} - 1 \mperiod
\end{align}
Repeating the above for $Y_\Sigma$ yields
\begin{align}
    \mathbb{E}_W\left[Y_\Sigma \mid \text{non-SCER Spoof}\right] &= 1 - \frac{2s}{n-r} + \frac{2sr}{(n-r)n} \mperiod
\end{align}

Conveniently, $\mathbb{E}_W\left[Y_\Delta \mid \text{non-SCER Spoof}\right] + \mathbb{E}_W\left[Y_\Sigma \mid \text{non-SCER Spoof}\right] = 0$, meaning that the sum of the two is not a function of $s$. 
Since the authentic provider knows the correct watermarked chips for every ranging code, the resulting expectations of the authentic case is $\mathbb{E}_W\left[Y_\Delta \mid \text{Authentic}\right] = \mathbb{E}_W\left[Y_\Sigma \mid \text{Authentic}\right] = 1$.
This results in $Y$ being zero mean in the spoofing case and having a mean of $2$ in the authentic case for each adversarial strategy $s$.

The variances under spoofing conditions are more involved, so we refer to \citet{anderson2024gnsscrypto}:
\begin{align}
    \mathbb{V}_W \left[Y_\Delta \mid \text{non-SCER Spoof} \right] &= \frac1W \frac 4r \frac sn \frac{n-r}{n} \frac{n-s}{n-1} + \frac 1W \frac 1r \frac{n}{FT} \frac{\sigma^2}{P} \msemi \label{eq: VD nSCER} \\
    \mathbb{V}_W \left[Y_\Sigma \mid \text{non-SCER Spoof} \right] &= \frac1W \frac{4}{n-r} \frac{sr}{n} \frac{1}{n} \frac{n-s}{n-1} + \frac 1W \frac{1}{n-r} \frac{n}{FT} \frac{\sigma^2}{P} \mperiod \label{eq: VS nSCER}
\end{align}
Since the authentic provider knows the correct watermarked chips for every ranging code, for the variances in the authentic case, the terms resulting from the hypergeometric distribution (i.e., the first terms in the sums of \Cref{eq: VD nSCER} and \Cref{eq: VS nSCER}) go away.

The covariance between $Y_\Delta$ and $Y_\Sigma$ is non-zero in the spoofing case, but negligible and not visible in any of our plots in the following sections.
Therefore, we do not provide here nor utilize in our experiments.
The variances ultimately determine whether the threshold decision boundary meets the security requirement.
If the variance is too large, then any decision threshold will not have the sensitivity and specificity requirement to be useful.

\section{Watermark Design}\label{sec:wm-design}

There are three primary factors when designing a watermarked ranging authentication protocol.

First, a watermark should provide a minimum specific security level.
In a cryptographic system, this usually refers to bounding the probability that the adversary will guess an underlying cryptographic secret or forging a cryptographic signature or authentication code.
Receivers do not examine individually observed ranging code chips; instead, they observe statistics that are related to the correlations against ranging code replicas.
This makes the adversarial forgery probability computation more involved.
In this context, the security level refers to bounding the probability that the adversary will be able to successfully select a ranging code that will fool the receiver's authentication statistics.
For the Pulsar X1 Watermark, the security requirement is 32-bits.
This means that the forgery probability must be less than $2^{-32}$.
This number was selected to conform with the security level of other GNSS constellation message authentication codes~\citep{dennis2024sbas}.
But those constellations selected those number to conform with other requirements so that the disruption to civil aviation is less than $10^{-7}$ or $10^{-9}$.
In all cases, this security level meets or exceeds the state of the art in GNSS authentication.

Second, the TTA should be as small as possible.
Pulsar X1 will transmit a navigation message each second.
To minimize the TTA, the watermark conforms to the X1 one-second message cadence.
This means that the Pulsar Watermark is designed to provide its security over a 1-second observation interval.
The remaining components of TTA related to the system-wide receiver time synchronization requirement and the time needed to transmit the watermark seed.

Third, a watermark should induce the minimum degradation required.
Satisfying this consideration ensures minimum disruption to Pulsar customers that wish to ignore the Pulsar authentication features and otherwise minimize the degradation to the Pulsar PNT capabilities.

These three concerns form a Pareto-Optimal design problem.
One can increase the watermark's security by increasing the number of chips inverted (thereby increasing the degradation).
Or one can increase the watermark's security by increasing the length of time of the watermark (thereby increasing the TTA).
And vice versa.

Following the procedure from \citet{anderson2024gnsscrypto}, we select the statistic and threshold of $Y = 1$ and utilize the CLT approximations of the radio processing statistics assuming $s=511$.
$s=511$ corresponds to the best adversarial strategy with the $Y=1$ decision boundary because that yields the maximum statistic variance under adversarial conditions (recall that the expectation is $0$ for all $s$).
Meeting a 32-bit security requirement over 1-second yielded a candidate of $(n=1023, r=21, W=1000)$ using the CLT approximations.
The following \Cref{sec:wm-sec} derives the exact probability and shows that $r=21$ is the minimum required inverted chips to minimize the watermark-induced degradation while maintaining a one-seconding interval and 32-bit security.

\section{Pulsar Watermark Security}\label{sec:wm-sec}

There are two relevant probabilities related to a watermark's security level.
First, we desire that the probability that an authenticate signal triggers an alarm to be small.
In the navigation context, this probability is called the Probability of False Alarm (PFA).
Second, we desire that the probability that an adversary can forge a watermark signal, according to our adversarial model, is also small.
In the navigation context, this probability is called the Probability of Missed Detection (PMD).
Since we have no reason to favor the PFA or the PMD, we set both to be the same as the security requirement from \Cref{sec:wm-design}:
\begin{align}
    \Pr \left( Y \leq 1 \mid \text{Authentic} \right) &< 2^{-32} \quad \label{eq:pfa-req} \msemi \\
    \Pr \left( Y > 1 \mid \text{Non-SCER Spoof}, s \right) &< 2^{-32} \quad \forall s \label{eq:pmd-req} \mperiod
\end{align}

Now we show that the Pulsar Watermark meets these two requirements.

\subsection{Probability of False Alarm}

From \citet{anderson2024gnsscrypto}, we have the distribution of $Y$ under authentic conditions:
\begin{align}
    Y \mid \textrm{Authentic} \sim \mathcal{N}\left(2, \frac1W \frac1r \frac{n}{FT} \frac{\sigma^2}{P} + \frac1W \frac{1}{n-r} \frac{n}{FT} \frac{\sigma^2}{P}\right) \mperiod \label{eq:Nauth}
\end{align}
Trivially, 
\begin{align}
    \Pr \left( Y \leq 1 \mid \text{Authentic} \right) &= \textrm{normcdf}_{N_{\Delta \Sigma W}} (-1) \msemi \\
    \Pr \left( Y \leq 1 \mid \text{Authentic} \right) &= 1.139 \cdot 10^{-10} < 2^{-32} \mperiod \nonumber
\end{align}

\subsection{Probability of Missed Detection}

Computing the Pulsar watermark PMD largely follows the procedures from \citet{anderson2024gnsscrypto}.
From \citet{anderson2024gnsscrypto}, we have the distribution of $Y_\Delta$ and $Y_\Sigma$ under non-SCER spoofing conditions:
\begin{align}
Y_\Delta \mid \textrm{Non-SCER Spoof} &= g_\Delta(H) + N_\Delta \msemi \tag{\ref{eq: YD nSCER}} \\
Y_\Sigma \mid \textrm{Non-SCER Spoof} &= g_\Sigma(H) + N_\Sigma \mperiod \tag{\ref{eq: YS nSCER}} 
\end{align}
Given our adversary model and security statistic, we now show that security of the Pulsar Watermark meets 32-bit security.
Recall that we must account for $Y$ under non-SCER spoofing conditions for any adversary strategy $s$.
Starting with \Cref{eq:pmd-req}, we condition on $s$ and decompose the underlying distributions into a convenient sum:
\begin{align}
    \Pr \left(Y > 1 \middle| s \right) =& \Pr \left(Y_\Delta + Y_\Sigma > 1 \middle| s \right) \mcomma \nonumber \\
    =& \Pr \left(\frac 1W \sum \left(g_\Delta(H) + N_\Delta \right) + \frac 1W \sum \left( g_\Sigma(H) + N_\Sigma \right) > 1 \middle| s \right) \mcomma \nonumber \msemi \\
    \Pr \left(Y > 1 \middle| s \right) =& \Pr \left(\frac 1W \sum g_\Delta(H) + \frac 1W \sum g_\Sigma(H) + \frac 1W \sum \left(N_\Delta + N_\Sigma\right) > 1 \middle| s \right) \mperiod
\end{align}
The normal terms amount to the average of the sum of two normal distributions, which is yet another normal distribution:
\begin{align}
    N_{\Delta \Sigma W} &\sim \mathcal{N}\left(0, \frac 1W \frac 1r \frac{n}{FT} \frac{\sigma^2}{P} + \frac 1W \frac{1}{n-r} \frac{n}{FT} \frac{\sigma^2}{P} \right) \msemi \label{eq:NDSW} \\
    \Pr \left(Y > 1 \middle| s \right) &= \Pr \left( \frac 1W \sum g_\Delta(H) + \frac 1W \sum g_\Sigma(H) + N_{\Delta \Sigma W} \geq 1 \middle| s \right) \mperiod \label{eq:before-conv}
\end{align}

In \Cref{eq:before-conv}, both terms involving $H$ can be computed directly via convolution because they are both known discrete probability distributions.
First, we assemble two arrays corresponding to the domain and probability mass function (PMF) for $H$.
This is the probability distribution of the adversary guessing the $r$ inverted chips among the $n$ total chips over a single watermark with $s$ inverted chip drawings.
Second, we compute the $W$-convolved $H$ probability distribution, as in \Cref{lst:HW}.
\begin{figure}
\centering
\begin{minipage}{0.5\linewidth}
\begin{lstlisting}[style=pseudopy, label={lst:HW}, caption={PMF Convolution Procedure}]
support_H = [0, ..., min(r, s)]
pmf_H = hypergeom.pmf(support_H, n, r, s)

support_HW = [0, ..., W * min(r, s)]
pmf_HW = []
for W times:
    pmf_HW = convolve(pmf_HW, pmf_H)
\end{lstlisting}
\end{minipage}
\end{figure}
\Cref{lst:HW} illustrates an inefficient method for simplicity.
A more efficient way to compute the convolved distribution is via repeated squaring.
The convolved distribution provides the probability distribution of the adversary guessing any chips correctly among the $W$ watermarks.
Third, to compute the convolved distributions transformed onto the domain of $Y$, we exploit the linearity of $g_\Delta$ and $g_\Sigma$ and follow the transformation rules for discrete probability distributions, as in \Cref{lst:gHW}.
This results in transforming the domains arrays without changing the PMF arrays.
\begin{figure}
\centering
\begin{minipage}{0.5\linewidth}
\begin{lstlisting}[style=pseudopy, label={lst:gHW}, caption={Final PMF Transformation Procedure}]
support_gdHW = gd(support_HW / W)
pmf_gdHW = pmf_HW
support_geHW = ge(support_HW / W)
pmf_geHW = pmf_HW
\end{lstlisting}
\end{minipage}
\end{figure}

Given \Cref{lst:HW,lst:gHW}, we now have two discrete distributions and a normal distribution, with the two discrete distributions being non-independent:
\begin{align}
    \Pr \left(Y > 1 \middle| s \right) &= \Pr \left( g_\Delta(H^{*W}) + g_\Sigma(H^{*W}) + N_{\Delta \Sigma W} > 1 \middle| s \right) \mperiod \label{eq:after-conv}
\end{align}
Now one could continue the convolution procedure; however, after incorporating $g_\Delta$ and $g_\Sigma$ the PMF domains no longer have integer support.
This prohibits the use of the Fast Fourier Transform, and, together with the need to discretize the normal distribution remaining, would not provide an exact answer.
Moreover, given that the arrays have length at most $rW$, convolving further requires enough memory and computational resources to necessitate a high performance computing cluster.

Conveniently, the convolved $H$ random variable captures all of the information connecting the two discrete distributions, which means all three distributions are conditionally independent on $H$:
\begin{align}
    \Pr \left(Y > 1 \middle| s \right) &= \sum_h \Pr \left( g_\Delta(H^{*W}) + g_\Sigma(H^{*W}) + N_{\Delta \Sigma W} > 1 \middle| h, s \right) \cdot \Pr(H^{*W} = h) \mcomma \\
    &= \sum_h \Pr \left( g_\Delta(h) + g_\Sigma(h) + N_{\Delta \Sigma W} > 1 \middle| s \right) \cdot \Pr(H^{*W} = h) \mcomma \nonumber \\
    &= \sum_h \Pr \left(N_{\Delta \Sigma W} > 1 - g_\Delta(h | s) - g_\Sigma(h | s) \right) \cdot \Pr(H^{*W} = h) \msemi \nonumber \\
    \Pr \left(Y > 1 \middle| s \right) &= \sum_h \left(1 - \textrm{normcdf}_{N_{\Delta \Sigma W}} ( 1 - g_\Delta(h | s) - g_\Sigma(h | s))\right) \cdot \Pr(H^{*W} = h) \mperiod \label{eq:pmd}
\end{align}

\begin{figure}
    \centering
    \includegraphics[width=0.5\linewidth]{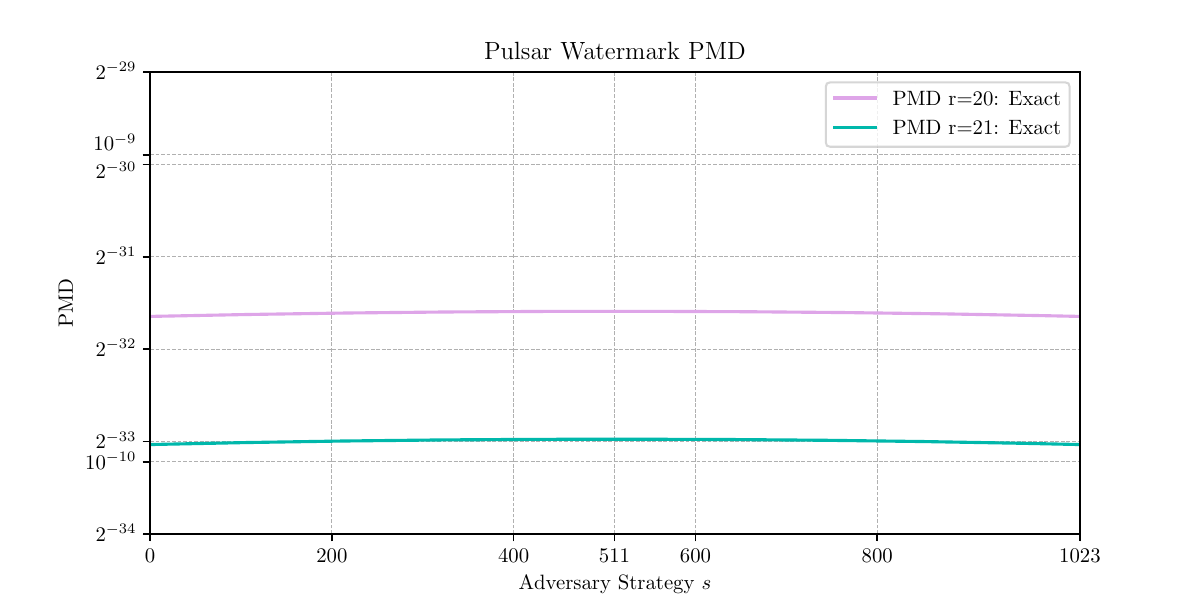}
    \includegraphics[width=0.5\linewidth]{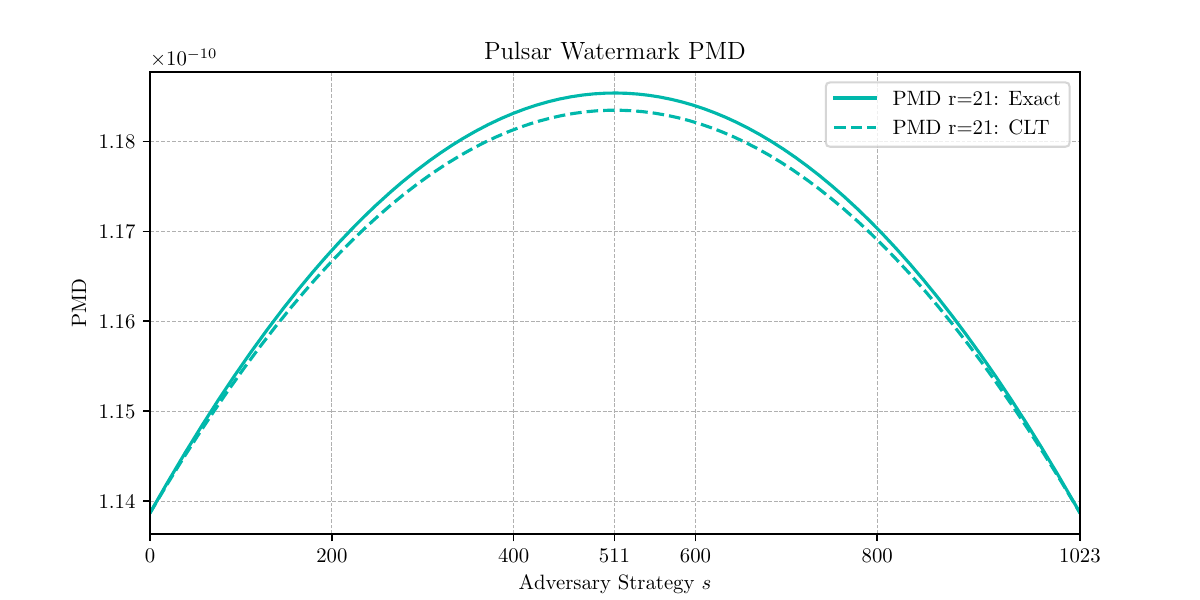}
    \caption{
        For both figures, the x axis is the adversarial strategy via their election $s$ and the y axis is the PMD.
        The top figure shows the PMD for a $r=20$ and a $r=21$ watermark, showing that $r=21$ is the smallest $r$ that meets 32-bit security.
        The bottom figure zooms into the $r=21$ line to compare the CLT approximation to the computed probability.
        }
    \label{fig:pmd}
\end{figure}

\Cref{fig:pmd} computes the PMD via \Cref{eq:pmd} over all $s$ to verify compliance with \Cref{eq:pmd-req} for each $s$-adversary strategy.
Because the PMD curve over $s$ for the $r=21$ watermark is less than the design requirement for every adversarial strategy $s$, the watermark meets the requirement.
In \Cref{fig:pmd}, we provide the PMD for the $r=20$ watermark to show that $r=21$ is the minimum required, and we compare to the CLT approximations initially used to determine $r=21$.

To compute \Cref{fig:pmd} in python, the convolved $H$ was computed via repeated squaring and the Fast Fourier Transform.
With 8 parallel workers, our laptops compute each PMD versus $s$ curve in about 10 seconds.

\section{Watermark Verification Results from Orbit} \label{sec:resultsFromOrbit}

On July 7th, 2025, Xona transmitted the first watermarked signal from orbit over Xona headquarters in Burlingame, CA.
Utilizing a recording, we perform the watermark verification methods from the previous sections after post processing with Xona's modified GNSS-SDR~\citep{gnsssdr}.
The recording exhibited $Y > 1$ for all tracked pseudoranges.

\begin{figure}
    \centering
    \includegraphics[width=0.47\linewidth]{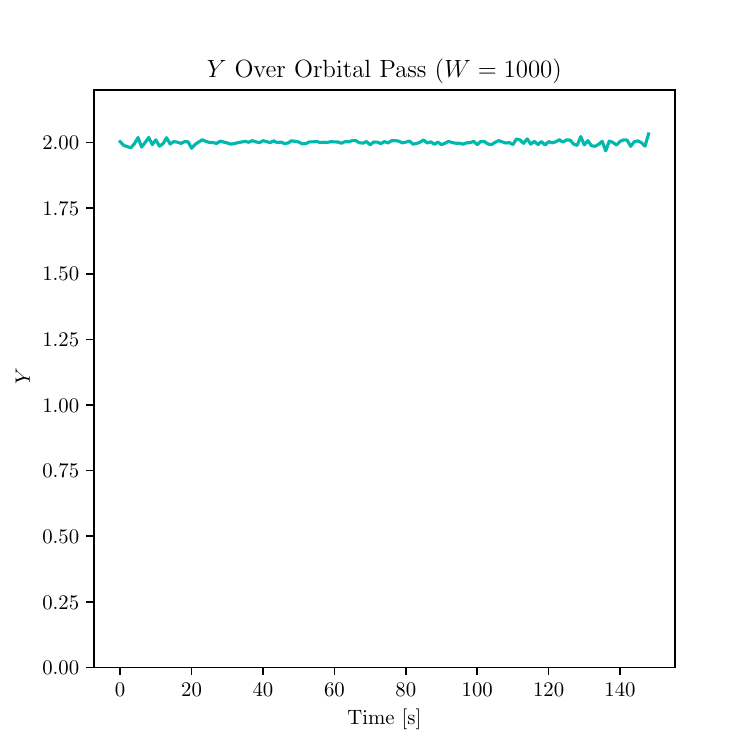}
    \includegraphics[width=0.47\linewidth]{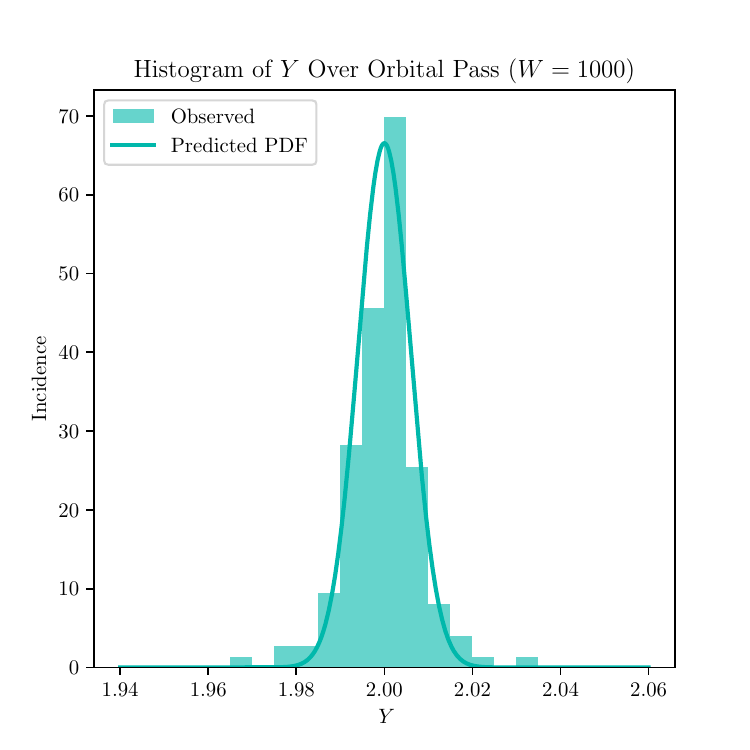}
    \caption{
        150 seconds of recorded data from a Pulsar-0 orbit pass over Xona headquarters in Burlingame, CA.
        The left plot provides the measured $Y$ over time for the Pulsar watermark ($W=1000$).
        The right plot provides a histogram of the same data points overlaid with the predicted distribution with the average $C/N_0$ over the pass.
        Over the pass, the $C/N_0$ changes from minimum to maximum and back to minimum which is observable in the left plot with the time-series magnitude.
        Because of these time-correlated artifacts, in the left plot, with respect to the predicted distribution using the average $C/N_0$, we expect more outlier from the beginning and end of the recording and greater concentration from the middle of the recording.
        }
    \label{fig:YoverTime}
\end{figure}

\Cref{fig:YoverTime} provides the measured $Y$ over the orbit pass over Xona's headquarters in Burlingame, CA.
In the left plot, the change in variance over time reflects the signal strength changing over the GNSS-SDR processing over the orbit pass.
In the right plot, the histogram compares the observed data with the predicted distribution utilizing the average $C/N_0$.
As expected, the histogram in the right plot of \Cref{fig:YoverTime} roughly matches the distribution predicted by \Cref{eq:Nauth}.

\section{Spoofing Experiment with Orbit Data} \label{sec:spoofExp}

\begin{figure}
    \centering
    \includegraphics[width=0.75\linewidth]{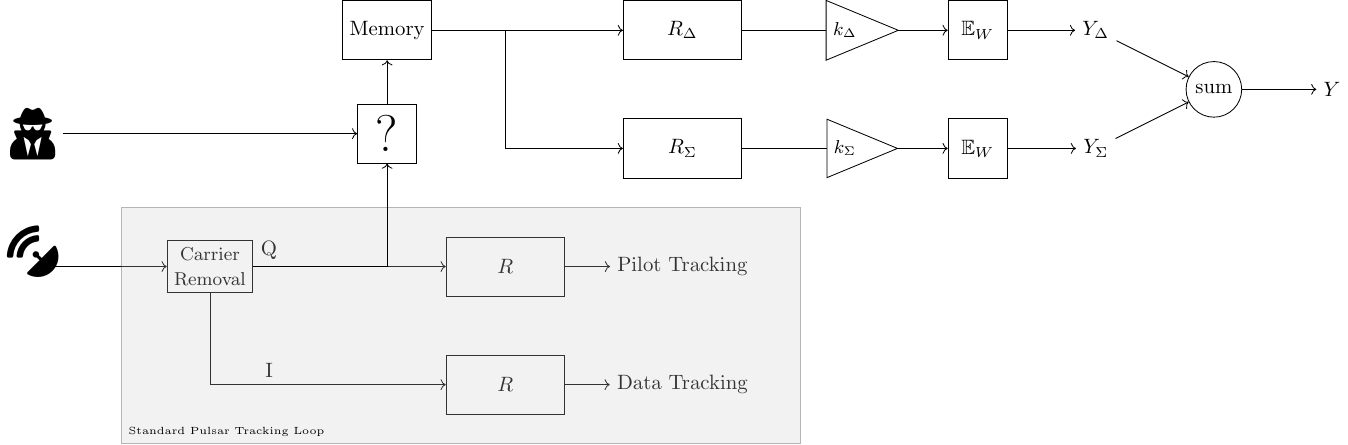}
    \caption{
        A conceptual diagram of our spoofing experiment where the watermark signal processing input is substituted by a spoofer-generated simulated signal.
        Simulated data with a varying number of spoofed inverted chips $s$ substitutes the Pulsar-0 recorded data.
        With \Cref{fig:spoof} we show that the Pulsar watermark and the statistics herein effectively enable the receiver to determine the authenticity of the signal.
        The left plot shows the results for the designed $W=1000$ Pulsar watermark.
        In the left plot, the markers themselves are larger than the 3-sigma ellipses of the predicted distribution, so we provide the right plot.
        The right plot provides results as if the Pulsar watermark were $W=50$ to provide a more interesting experiment that demonstrates our ability to predict the spread of the distributions.
        The ellipses are the 3-sigma confidence assuming the minimum $C/N_0$ over the orbit pass, and the spoofing simulations were simulated assuming the average $C/N_0$ over the orbit pass.
        Therefore, we expect both the simulated spoofer results and the orbit recording results to be more concentrated that the provided ellipses. 
        }
    \label{fig:concept-experiment}
    \end{figure}

For the avoidance of doubt, we develop a quick spoofing experiment to demonstrate the efficacy of Pulsar's watermark.
\Cref{fig:concept-experiment} provides a conceptual diagram of the experiment.
The tracking loop is converged with the orbit data, and then a spoofing signal is injected after the GNSS-SDR phase and doppler rotation correction.
Each spoofed signal is a signal with the same estimated SNR, but the watermarked chips are randomly generated.
The spoofed signals have varying adversary-selected number of chips inverted $s$.
For \Cref{fig:concept-experiment}, the simulated adversaries include $s=0, 200, 400, 600, 800, 1023$.

For the case of $s=1023$, every chip is inverted.
This would be as if the spoofer submitted an unwatermarked ranging code with inverted phase to the receiver.
Because there is an instantaneous change from the authentic orbit data to the spoofed injection, the receiver still immediately rejects the signal.
There is no time for the tracker to attempt to invert the phase.
Were this $s=1023$ attack to continue long term, the tracking loop would invert the phase (so as to appear as an $s=0$ spoofing attack) before rejecting the signal long term.

\begin{figure}
    \centering
    \includegraphics[width=0.47\linewidth]{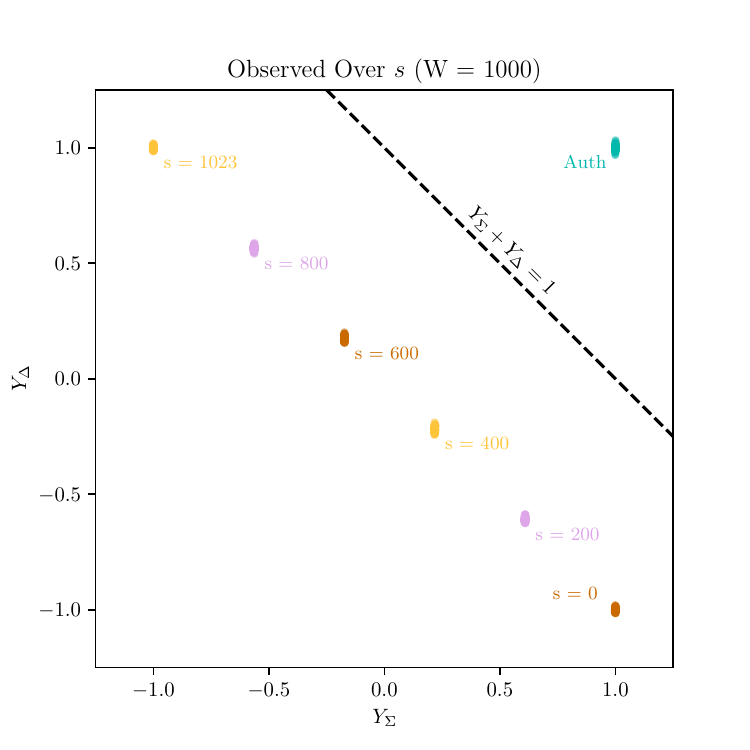}
    \includegraphics[width=0.47\linewidth]{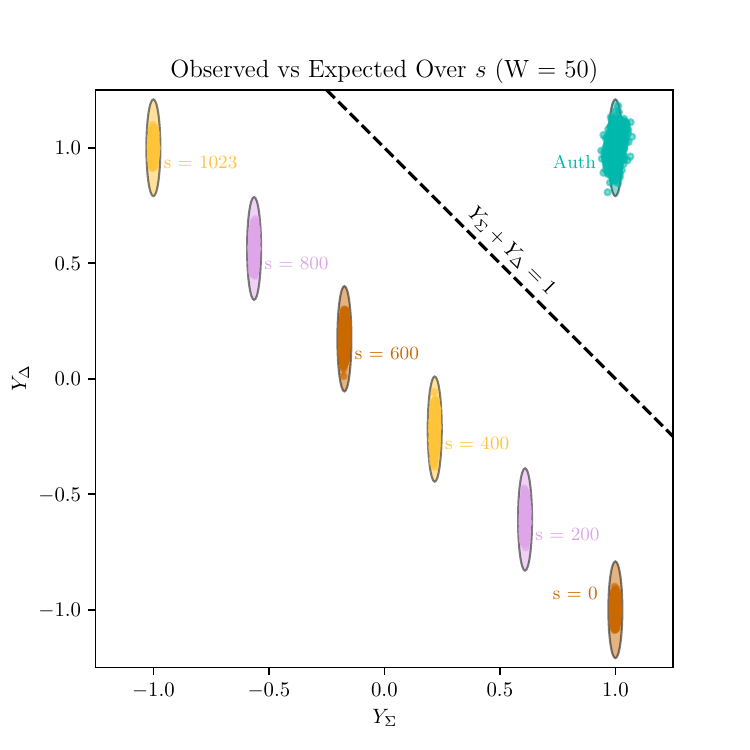}
    \caption{
        Results from the spoofing experiment with the orbit data.
        On the left are the results where the receiver is complying with the specified $W=1000$ Pulsar watermark.
        Because the large $W=1000$ average results in the markers exceeding the size of the predicted $3\sigma$ ellipses, we provide the right plot.
        The right plot is with $W=50$, meaning the receiver is aggregating 50 ms of data to determine authenticity.
        This induces a noisy result that is not compliant with the 32-bit security requirement but provides a more interesting plot.
        $s$ is the number of chips the adversary randomly flips in the simulated signal that substitutes the data recorded from Pulsar-0.
        In the right plot, the ellipses correspond to the worst $C/N_0$ over the orbit pass; hence, the data points are more concentrated than the provided ellipses.
        }
    \label{fig:spoof}
\end{figure}

\Cref{fig:spoof} provides the results of the experiment diagramed by \Cref{fig:concept-experiment}.
Each data point represents a coherent observation of a $(Y_\Delta, Y_\Sigma)$ pair after a $W$ average.
The left plot corresponds to the actual Pulsar watermark, requiring the average of $W=1000$.
Because the Pulsar watermark security is set to $2^{-32}$ with $W=1000$, the width of the markers is wider than the predicted 3$\sigma$ ellipses of the CLT approximations, making an uninteresting figure.
Hence, we provide the right plot which corresponds to a watermark with $W=50$, which better demonstrates the efficacy of the mathematical predictions in this work.

The left plot in \Cref{fig:spoof} demonstrates the effect of having 32-bit security and operating at a $C/N_0$ well above 30 dB-Hz.
It is extremely unlikely for the spoofing statistic to fail.
The right plot demonstrates how we can predict the center and spread of the statistical distributions under authentic and spoofing conditions.

\section{Conclusion}

With this work, we verify the authenticity of a signal from Pulsar-0 utilizing the Pulsar Combinatorial Watermark.
With the derivations herein, we show that the Pulsar Watermark provides at least 32-bit security.
Because of (1) the cryptographic construction lacking any assumptions about the possession of symmetric keys, (2) we show how to mathematically derive a bound on the PMD, and (3) the orbit results presented in this work, we claim the world's first authenticated satellite pseudorange from orbit.
With the experiments herein, we show that the mathematical predictions match the experimental data from orbit.
Moreover, we demonstrate the Pulsar Watermark's efficacy against injected spoofing signals.

\bibliographystyle{apalike}
\bibliography{bibliography}

\end{document}